\documentclass[aps,prc,amsmath,amssymb,superscriptaddress,noshowpacs,showkeys]{revtex4}
\usepackage{dcolumn}
\usepackage{graphicx,color,subfigure}
\usepackage{multirow}
\usepackage{textcomp}
\usepackage{relsize}
\usepackage{tikz}	
\usetikzlibrary
{
	arrows,
	calc,
	through,
	shapes.misc,	
	shapes.arrows,
	chains,
	matrix,
	intersections,
	positioning,	
	scopes,
	mindmap,
	shadings,
	shapes,
	backgrounds,
	decorations.text,
	decorations.markings,
	decorations.pathmorphing,	
}
\begin {document}
  \newcommand {\nc} {\newcommand}
  \nc {\beq} {\begin{eqnarray}}
  \nc {\eeq} {\nonumber \end{eqnarray}}
  \nc {\eeqn}[1] {\label {#1} \end{eqnarray}}
  \nc {\eol} {\nonumber \\}
  \nc {\eoln}[1] {\label {#1} \\}
  \nc {\ve} [1] {\mbox{\boldmath $#1$}}
  \nc {\ves} [1] {\mbox{\boldmath ${\scriptstyle #1}$}}
  \nc {\mrm} [1] {\mathrm{#1}}
  \nc {\half} {\mbox{$\frac{1}{2}$}}
  \nc {\thal} {\mbox{$\frac{3}{2}$}}
  \nc {\fial} {\mbox{$\frac{5}{2}$}}
  \nc {\la} {\mbox{$\langle$}}
  \nc {\ra} {\mbox{$\rangle$}}
  \nc {\etal} {\emph{et al.}}
  \nc {\eq} [1] {(\ref{#1})}
  \nc {\Eq} [1] {Eq.~(\ref{#1})}
  \nc {\Ref} [1] {Ref.~\cite{#1}}
  \nc {\Refc} [2] {Refs.~\cite[#1]{#2}}
  \nc {\Sec} [1] {Sec.~\ref{#1}}
  \nc {\chap} [1] {Chapter~\ref{#1}}
  \nc {\anx} [1] {Appendix~\ref{#1}}
  \nc {\tbl} [1] {Table~\ref{#1}}
  \nc {\fig} [1] {Fig.~\ref{#1}}
  \nc {\ex} [1] {$^{#1}$}
  \nc {\Sch} {Schr\"odinger }
  \nc {\flim} [2] {\mathop{\longrightarrow}\limits_{{#1}\rightarrow{#2}}}
  \nc {\textdegr}{$^{\circ}$}
  \nc {\inred} [1]{\textcolor{red}{#1}}
  \nc {\inblue} [1]{\textcolor{blue}{#1}}
  \nc {\IR} [1]{\textcolor{red}{#1}}
  \nc {\IB} [1]{\textcolor{blue}{#1}}
  \nc{\pderiv}[2]{\cfrac{\partial #1}{\partial #2}}
  \nc{\deriv}[2]{\cfrac{d#1}{d#2}}
\title{Study of corrections to the eikonal approximation \\[0.2cm] {\small Contribution to the 55$^\mathrm{th}$ International Winter Meeting on Nuclear Physics (January 2017, Bormio, Italy)}}
\author{C.~Hebborn}
\email[FRIA grant. Electronic address: ]{chloe.hebborn@ulb.ac.be}
\affiliation{Physique Nucl\' eaire et Physique Quantique (CP 229), Universit\'e libre de Bruxelles (ULB), B-1050 Brussels, Belgium}
\author{P.~Capel\textsuperscript{1,}}
\email{pierre.capel@ulb.ac.be}
\affiliation{Institut f\"ur Kernphysik, Technische Universit\"at Darmstadt, D-64289 Darmstadt, Germany }
\date{\today}
\begin{abstract}
	For the last decades, multiple international facilities have developed Radioactive-Ion Beams (RIB) to measure reaction processes including exotic nuclei. These measurements coupled with an accurate theoretical model of the reaction enable us to infer information about the structure of these nuclei. The partial-wave expansion  provides a precise description of two-body collisions but has a large computational cost. To cope with this issue, the eikonal approximation is a powerful tool as it reduces the computational time and still describes the quantum effects observed in reaction observables. However, its range of validity is restricted to high energy and to forward scattering angles.  In this work, we analyse the extension of the eikonal approximation to lower energies and larger angles through the implementation of two corrections. These aim to improve the treatment of the nuclear and Coulomb interactions within the eikonal model. The first correction is based on an expansion of the $T$-matrix while the second relies on a semi-classical approach. They permit to better account for the deflection of the projectile by the target, which is neglected in the standard eikonal model.   The gain in accuracy of each correction is evaluated through the analyses of angular cross sections computed with the standard eikonal model, its corrections and the partial-wave expansion. These analyses have been performed for tightly bound projectiles ($^{10}\mathrm{Be}$) from intermediate energies (67~MeV/nucleon) down to energies of interest of future RIB facilities such as HIE-ISOLDE and ReA12 at MSU (10~MeV/nucleon).  
\end{abstract}

\maketitle
%


\section{Introduction}\label{Introduction}

The development of Radioactive-Ion Beams (RIB) has enabled the discovery of nuclei with very unexpected structures. In particular, in the neutron-rich region of the nuclear chart, halo nuclei have been observed. These exotic nuclei exhibit a very large matter radius due to the low binding energy of one or two neutrons, which allows them to decouple from the core of the nucleus and to form a diffuse halo~\cite{Tanihata1996,Pietro2010}. They are modelled as two- or three-body objects: a compact core and one or two valence neutrons. As they are very short-lived, they cannot be studied through usual spectroscopic techniques but we can infer information about their structure from measurements of reaction processes coupled with an accurate model of reaction~\cite{BayeCapel,AlkhaliliNunes2003}.

Nowadays some RIB facilities, like ISOLDE at CERN, provide exotic beams at 5 MeV/nucleon and the goal is to reach 10~MeV/nucleon. At such energies, precise models such as the Continuum-Discretised Coupled Channel method (CDCC, see Refs.~\cite{BayeCapel,AlkhaliliNunes2003,ThompsonNunes2009})  have large computational cost and can present convergence problems. The eikonal approximation is a  quantal method which has a reduced computational time and can be easily interpreted. Unfortunately, it is valid only at high energies~\cite{Glauber}. In this work, we investigate the extension of this model to lower energies by the study of two corrections.

These corrections have already given interesting results for different types of reaction at various energies. Indeed, the first correction, proposed by Wallace (see Refs.~\cite{Wallace}), has also been analysed in Refs.~\cite{AlKhaliliTostevin,AguiarZardiVitturi1997,BuuckMiller,OvermeireRyckebusch}. Moreover, Refs.~\cite{AguiarZardiVitturi1997,LenziVitturiZardi1995,FukuiOgataCapel2014} have demonstrated the efficiency of the second correction for Coulomb-dominated collisions. Since we aim to describe light halo nuclei, our research focuses on nuclear dominated reactions. In this work, we provide analyses  of these corrections to simple cases, two-body collisions. Our final goal is to generalise them to three- and four-body collisions.  

In \Sec{TheoreticalFramework}, we describe the eikonal approximation for the elastic scattering and the two aforementioned corrections. Then, in \Sec{Results}, the numerical inputs used in our computations are given. 
From the analyses conducted on the differential cross sections for the elastic scattering of  \ex{10}Be off \ex{12}C, we conclude and propose an idea to pursue the extension of the eikonal model.

\section{Theoretical framework}\label{TheoreticalFramework}
\subsection{The eikonal description of elastic scattering}\label{eikonal}
In this study, we consider the elastic scattering of a projectile $P$ of mass $m_P$ and charge $Z_P e$  impinging on a target $T$ of mass $m_T$ and charge $Z_T e$. We assume both nuclei to be structureless and spinless and their interaction to be simulated by an optical potential $V$.
Their relative motion can hence be described by the following Schr\"odinger equation
\begin{equation}
\left[\frac{P^2}{2\mu} + V(\ve{R})\right] \Psi(\ve{R}) = E\Psi(\ve{R}),
\label{EqSchrodinger}
\end{equation}
where $\ve{R}$ is the projectile-target relative position, $\ve{P}$ the corresponding momentum, $\mu = m_P m_T/(m_P+m_T)$ the $P$-$T$ reduced mass and $E$ the total energy of the system in the center-of-mass restframe.
For the following developments, we assume the potential to be central.

\begin{figure}
	\centering
	\includegraphics[width=0.3\linewidth]{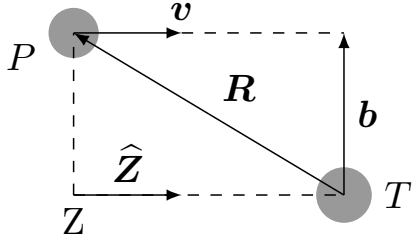}
	\caption{\label{FigEIkCoordinates} Coordinate system: the projectile-target relative coordinate $\ve{R}$ expanded in its transverse $\ve{b}$ and longitudinal $Z$ components.}
\end{figure}

To describe the aforementioned collision, \Eq{EqSchrodinger} has to be solved with the condition that the projectile is impinging on the target with an initial momentum $\hslash \ve{K}=\hslash K \ve{\widehat Z}$, i.e., whose direction we choose for the $Z$ axis (see \fig{FigEIkCoordinates})
\beq
\Psi(\ve{R})\flim{Z}{-\infty}e^{iKZ+\cdots}.
\eeqn{eCL}
The ``$\cdots$'' in this asymptotic expression indicates that the incoming plane wave is distorted by $V$, even at large distances.

The eikonal approximation assumes that at sufficiently high energy, the relative motion of the nuclei does not differ much from the initial plane wave of \Eq{eCL}.
It is then suggested to factorize that plane wave out of the wave function $\Psi$ \cite{BayeCapel,Glauber,Bertulani}
\begin{equation}
\Psi(\ve{R})=e^{i K Z}\ \widehat{\Psi}(\ve{R}).
\label{EqEikonalFactorization}
\end{equation}
The new wave function $\widehat{\Psi}$, depending smoothly on $\ve{R}$, enables us to simplify the \Sch equation \eq{EqSchrodinger}.
Inserting \Eq{EqEikonalFactorization} into \Eq{EqSchrodinger} and neglecting the second-order derivative of $\widehat \Psi$ in comparison to its first-order derivative, leads to \cite{BayeCapel,Glauber,Bertulani}
\begin{equation}
i\hslash v \pderiv{}{Z} \widehat{\Psi}(\ve{b},Z) =V(b,Z)\widehat{\Psi}(\ve{b},Z),
\label{EqSchrodingerEikonal}
\end{equation} 
where $v=\hslash K/\mu$ is the initial velocity of the projectile relative to the target.
In \Eq{EqSchrodingerEikonal}, we express explicitly the dependence of $\widehat \Psi$ on the transverse $\ve{b}$ and longitudinal $Z$ components of $\ve{R}$ as illustrated in \fig{FigEIkCoordinates}.

The solutions of \Eq{EqSchrodingerEikonal} read \cite{BayeCapel,Glauber,Bertulani}
\begin{equation}
\widehat{\Psi}(\ve{b},Z)=\exp\left[-\frac{i}{\hslash v} \int^{Z}_{-\infty}{V(\ve{b},Z')\mathrm{d}Z'}\right].
\end{equation}
This eikonal approximation of the wave function has a simple semiclassical interpretation: the projectile is seen as moving on a straight-line trajectory, accumulating a complex phase through its interaction with the target.

The scattering amplitude can be derived from these solutions as \cite{BayeCapel,Glauber,Bertulani}
\begin{equation}
f(\theta)=-\frac{iK}{2\pi} \int \mathrm{d^2}b \left\{\exp \left[i\chi_0 (b)\right]-1\right\}\exp \left(i\ve{q} \cdot \ve{b}\right),
\label{e6} 
\end{equation}
where
\begin{equation}
\chi_0(b) = -\frac{1}{\hslash v} {\int_{-\infty}^{\infty} V(b,Z) dZ} \label{e7}
\end{equation}
is the eikonal phase and $\hslash\ve{q}= \hslash\ve{K'}-\hslash K \ve{\widehat Z}$ the momentum transferred during the scattering process to reach the final momentum $\hslash\ve{K'}$.

Since the eikonal phase \Eq{e7} diverges for the Coulomb potential, that interaction requires a particular treatment.
As indicated in \Ref{Bertulani}, one should simply add to the eikonal phase computed with the nuclear part of the optical potential the Coulomb eikonal phase
\begin{equation}
\chi^C=2\eta \ln(Kb), \label{e8}
\end{equation}
where $\eta=Z_PZ_T e^2/4\pi\epsilon_0\hslash v$ is the Sommerfeld parameter.
That phase leads to the exact Coulomb scattering amplitude.

At low energy, the eikonal approximation is no longer valid.
However, since its implementation and interpretation are straightforward, it would be useful to extend its domain of validity to low energy.
In the present work, we study two corrections which aim to take into account the deflection of the projectile by the target.
The first one, proposed by Wallace (see Refs.~\cite{Wallace}), acts on the nuclear interaction, while the second one can be applied to both interactions \cite{AguiarZardiVitturi1997,LenziVitturiZardi1995}.

\subsection{Wallace's correction}\label{CorrWallace}
It is derived from an expansion of the $T$-matrix whose exact form reads \cite{Wallace,Bertulani}
\begin{equation}
T = V+VGV, \label{EqTransitionMatrix}
\end{equation} 
where $G$ is the exact propagator defined by $G^{-1} = E-P^2/2\mu - V +i\varepsilon$.
This propagator can be expressed in terms of the eikonal propagator $g$ and a corrective term $N$ accounting for the deviations of the wave vector from the average wave vector $\widetilde{\ve{K}}=(\ve{K'}+\ve{K})/2$ due to the $P$-$T$ interaction during the reaction process  \cite{Wallace}
\begin{eqnarray}
G&=&g+gNG. \label{EqExactPropagatorEikonalPropagator}
\end{eqnarray}
The eikonal propagator can be derived by expanding the momentum $\ve{P}$ around the average wave vector and neglecting the quadratic terms 
\beq
g^{-1}&=&\widetilde{\ve{v}} \cdot (\widetilde{\ve{K}}-\ve{P}) -V +i\varepsilon,\label{EqEikonalPropagator}
\eeq
and its correction is defined by
\beq
N&=& \left(1-\cos{\frac{\theta}{2}}\right) (g^{-1} +V)+\frac{\hslash^2}{2\mu}\left[(\ve{P}-\ve{K'})\cdot(\ve{P}-\ve{K})\right], \label{EqCorrectionN}
\end{eqnarray} 
where $\widetilde{\ve{v}}=\hslash \widetilde{\ve{K}}/\mu$ is the average velocity.

Wallace has obtained an expansion of the $T$-matrix by inserting iteratively the relation \Eq{EqExactPropagatorEikonalPropagator} into \Eq{EqTransitionMatrix} \cite{Wallace}
\begin{eqnarray}
T&=&(V+VgV)+VgNgV +...,
\end{eqnarray}
where the terms in parenthesis correspond to the standard eikonal approximation while the others are corrections.

In Refs.~\cite{Wallace}, it is shown that, due to the rotational invariance of the potential, the correction terms depending explicitly on the scattering angle $\theta$ cancel. Therefore, the scattering amplitude at the $m^\mathrm{th}$ order can be expressed as 
\beq
f^{(m)}(\theta)&=&-\frac{iK}{2\pi} \int \mathrm{d^2}b \mathcal{T}^{(m)}(b)\exp \left(i\ve{q} \cdot \ve{b}\right).
\eeqn{e14}
The zeroth order $\mathcal{T}^{(0)}$ corresponds to the standard eikonal model, developed by Glauber~\cite{Glauber} [see \Eq{e6}]. Wallace has derived the first three corrected orders of  the $T$-matrices $\mathcal{T}^{(m)}$
\beq
\mathcal{T}^{(\mathrm{I})}(b)&=& \exp \left\{i\left[\chi_0 (b)+\tau_1 (b)\right]\right\}-1 \\
\mathcal{T}^{(\mathrm{II})}(b)&=& \exp \left\{i\left[\chi_0 (b)+\tau_1 (b)+\tau_2(b)\right]\right\}  \exp \left[-\omega_2 (b)\right]-1 \\
\mathcal{T}^{(\mathrm{III})}(b)&=& \exp \left\{i\left[\chi_0 (b)+\tau_1 (b)+\tau_2(b)+\tau_3(b)+\phi_3(b)\right]\right\} \exp\left\{-\left[\omega_2 (b)+\omega_3(b)\right]\right\}-1  \label{EqTransitionMatrixOrderM},
\end{eqnarray}
with the additional phases defined as
\beq
\tau_1 (b)&=& -\frac{\epsilon}{\hslash v}\,  (1+\beta_1)\int^\infty_{0} V^2(R) \mathrm{d}Z \label{EqTau1} \\
\tau_2(b)&=&-\frac{\epsilon^2}{\hslash v} \,  \left(1+\frac{5}{3}\beta_1 +\frac{1}{3}\beta_2\right)\int^\infty_{0} V^3(R) \mathrm{d}Z- \frac{b}{24K^2} \left[\chi_0' (b)\right]^2 \label{EqTau2} \\
\tau_3(b) &=&-\frac{\epsilon^3}{\hslash v} \, \left(\frac{5}{4} +\frac{11}{4}\beta_1 +\beta_2 +\frac{1}{12}\beta_3\right) \int^\infty_{0}  V^4(R) \mathrm{d}Z -\frac{b}{8K^2}  \tau_1'(b) \left[\chi_0'(b)\right]^2  \label{EqTau3}\\
\phi_3(b) &=&  -\frac{\epsilon}{\hslash v} \, \left( 1+\frac{5}{3} \beta_1+\frac{1}{3} \beta_2\right) \int^\infty_{0} \left[ \frac{1}{2K} \pderiv{V}{R}(R)\right]^2 \mathrm{d}Z \label{EqPhi3} \\
\omega_2(b) &=& \frac{b}{8K^2} \chi_0'(b) \nabla^2\chi_0(b) \label{EqOmega2} \\
\omega_3(b) &=&\frac{1}{8K^2} \left[b \chi_0'(b) \nabla^2 \tau_1(b) +b \tau_1'(b) \nabla^2\chi_0(b) \right].  \label{EqOmega3} 
\end{eqnarray}
where $\beta_n \equiv b^n \pderiv{^n}{b^n}$ is the transverse derivatives and $\epsilon =1/E$ the expansion parameter~\cite{Wallace}.

Two main limitations of this correction are identified in Refs.~\cite{Wallace}.
The expansion suggested by Wallace is valid only at sufficiently large energies and for potentials which vary smoothly.
These conditions ensure that the expansion parameter $\epsilon$ and the derivatives of the potential take small values, which is necessary for the perturbation treatment to hold. 
Physically, it is due to the fact that, at lower energies, the $P$-$T$ relative motion differs more from the initial plane wave [see \Eq{EqEikonalFactorization}]. 
We can also note that this correction is only significant for the nuclear interaction since the corrective terms vanish exactly for a potential decreasing as $1/r$.

\subsection{Semi-classical correction}\label{CorrSC}

As mentioned in \Sec{eikonal}, within the eikonal approximation framework, the projectile is seen as moving along straight-line trajectories and therefore the deflection of the projectile by the target is neglected. 
To improve the simulation of the Coulomb interaction within the eikonal model, we can replace the actual impact parameter $b$ in the eikonal phase by the distance of closest approach between the projectile and the target in the corresponding Coulomb trajectory \cite{Bertulani,BrogliaWinther}
\begin{equation}
\chi(b) \rightarrow \chi(b') \quad \text{with} \quad b'=	 \frac{\eta + \sqrt{\eta^2 +b^2 K^2}}{K}.\label{EqCorrSCC}
\end{equation}
In \Ref{FukuiOgataCapel2014} it was observed that this correction enables to account efficiently for that deflection in the Coulomb breakup of halo nuclei.

Similarly, an extension of this correction to the nuclear interaction is used in Refs.~\cite{AguiarZardiVitturi1997,LenziVitturiZardi1995}. To also account for the deflection due to the nuclear interaction, they have proposed to apply the same idea with the distance of closest approach $b'$ between the two nuclei of the classical trajectory considering both interactions~\cite{AguiarZardiVitturi1997,LenziVitturiZardi1995}
\begin{equation}
\chi(b) \rightarrow \frac{b'}{b}\chi(b').\label{EqCorrSCNC}
\end{equation}
The ratio $b'/b$ ensures the conservation of the angular momentum. It is equivalent to change the asymptotic velocity by the tangential velocity at the distance of closest approach. In this study, $b'$ is calculated from  the real part of the potentials, i.e. the Coulomb potential and the real part of the nuclear optical potential.

\section{Results}\label{Results}
\subsection{Projectile-target potentials}
To analyse the effects of the corrections presented in Sec.~\ref{TheoreticalFramework}, we consider the elastic scattering of a nuclear-dominated reaction ($^{10}\mathrm{Be}$ off $^{12}\mathrm{C}$) at different energies.
In this work, we use the potential developed in~\cite{AlKhaliliTostevin} to simulate the \ex{10}Be-\ex{12}C interaction. The nuclear part is given by
\begin{eqnarray}
V_N (R) &=& -V_R f_{WS}(R,R_R,a_R) -iW_I f_{WS}(R,R_I,a_I) -i  4 a_D W_D \deriv{f_{WS} }{R} (R,R_D,a_D), \label{EqWSPot} \\
\text{with}&& \quad f_{WS}(R,R_X,a_X) = \frac{1}{1 +e^{\frac{R-R_X}{a_X}}},
\label{WSFormFactor}
\end{eqnarray}
with the different parameters listed in Table~\ref{Tab10Be12CPotentialParameter}.
\begin{table}
	\center
	\begin{tabular}{ccc}
		\hline\hline
		$V_R = 123.0$ MeV & $R_R = 3.33$ fm & $a_R=  0.80$ fm \\
		$W_I = 65.0$ MeV & $R_I = 3.47$ fm & $a_I=0.80$ fm\\
		\hline\hline
	\end{tabular}
	\caption{Parameters of the potential used to simulate the $^{10}\mathrm{Be}$-$^{12}\mathrm{C}$ nuclear interaction [see \Eq{EqWSPot}, $W_D=0$ here]. This potential is taken from \Ref{AlKhaliliTostevin}.}
	\label{Tab10Be12CPotentialParameter}
\end{table}
The Coulomb part of the interaction is simulated by the potential of an uniformly charged sphere of radius $R_C=1.2\times (10^{1/3}+12^{1/3})$~fm \cite{AlKhaliliTostevin}.
Since the goal of the present study is to compare the standard eikonal model with its corrections, we use the same potential for all calculations and we neglect any energy dependence.

\subsection{Analysis}
To evaluate the gain brought by each corrections presented in \Sec{TheoreticalFramework}, we compare the cross sections for elastic scattering and the $T$-matrices at two energies (67 and 10~MeV/nucleon) with results obtained with a fully-converged partial-wave expansion, considered as exact. 

\begin{figure*}[t]
	\subfigure[\ ]{
		{\includegraphics[width=0.45\linewidth]{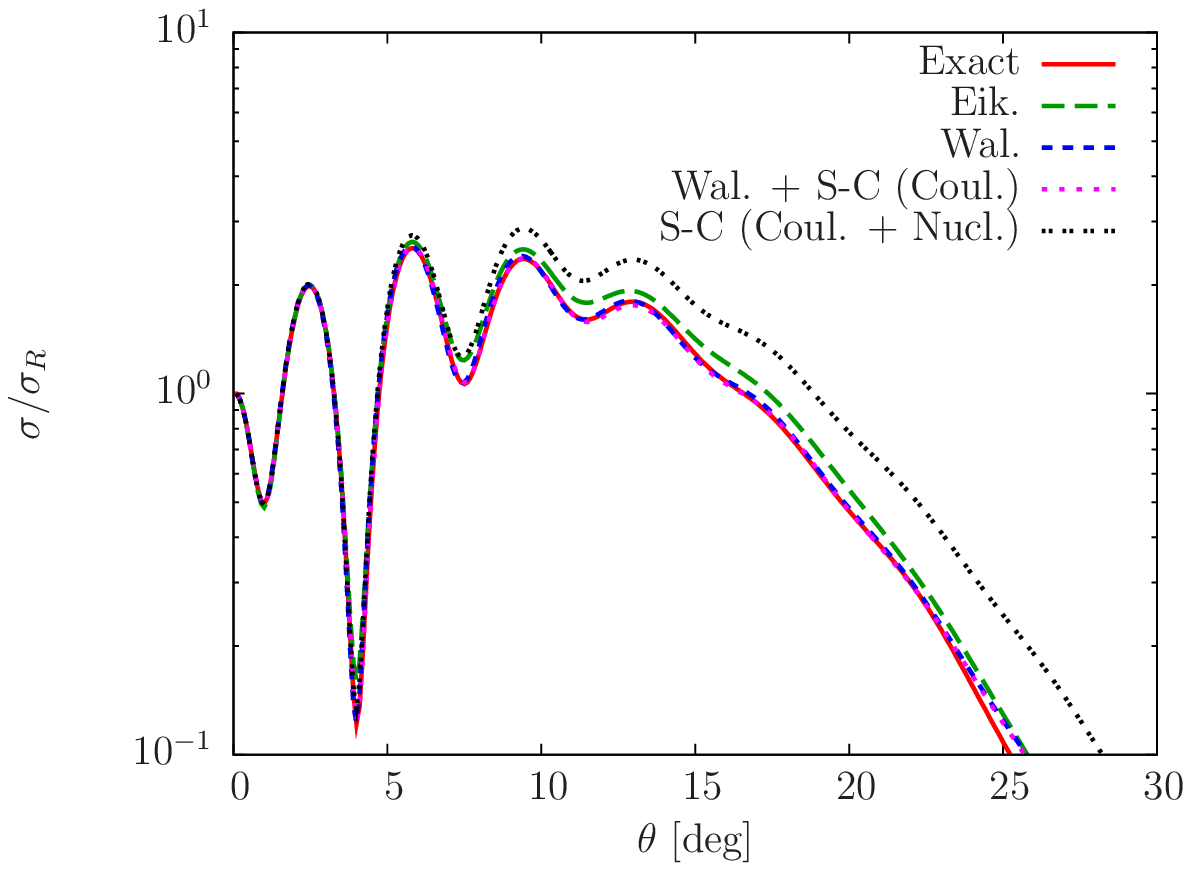}}
	}
	\subfigure[\ ]{	
{		\includegraphics[width=0.485\linewidth]{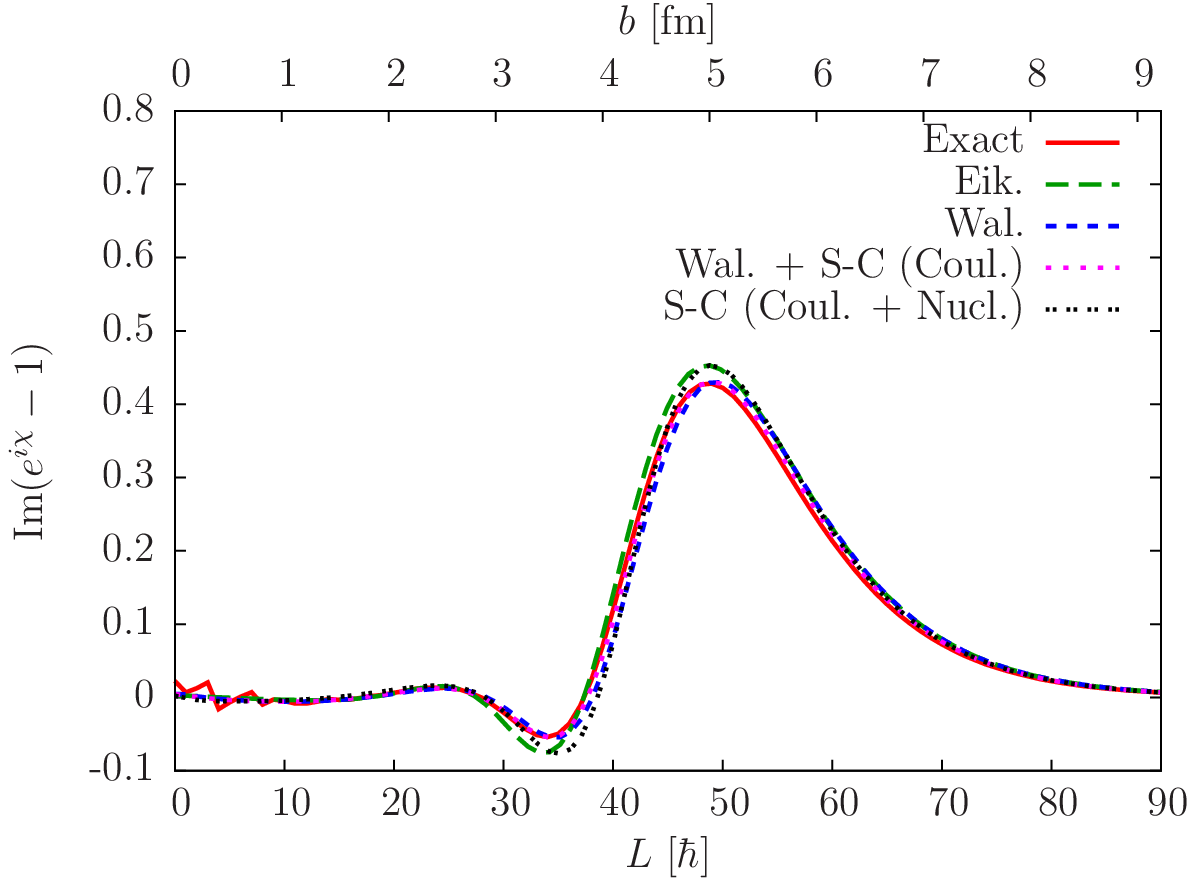}
	}}
\subfigure[\ ]{	
{		\includegraphics[width=0.45\linewidth]{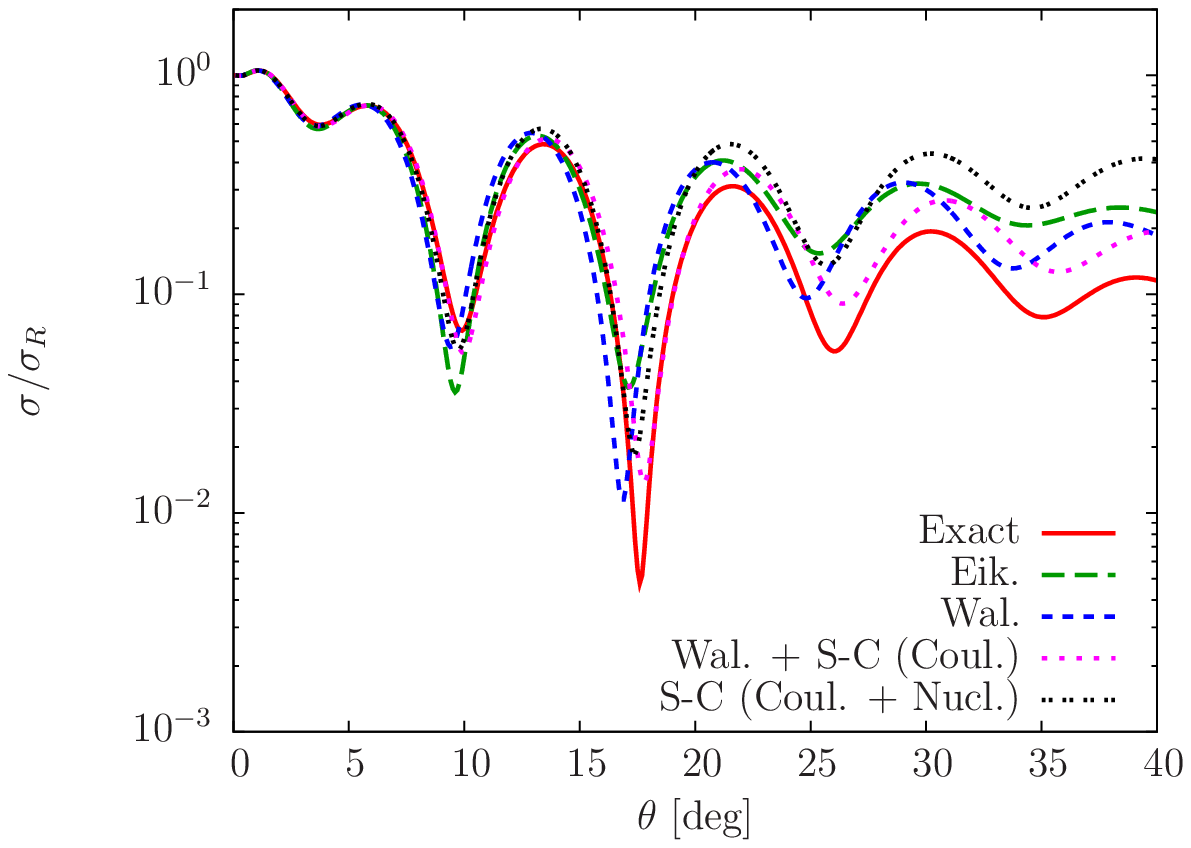}
	}}
\subfigure[\ ]{
{		\includegraphics[width=0.485\linewidth]{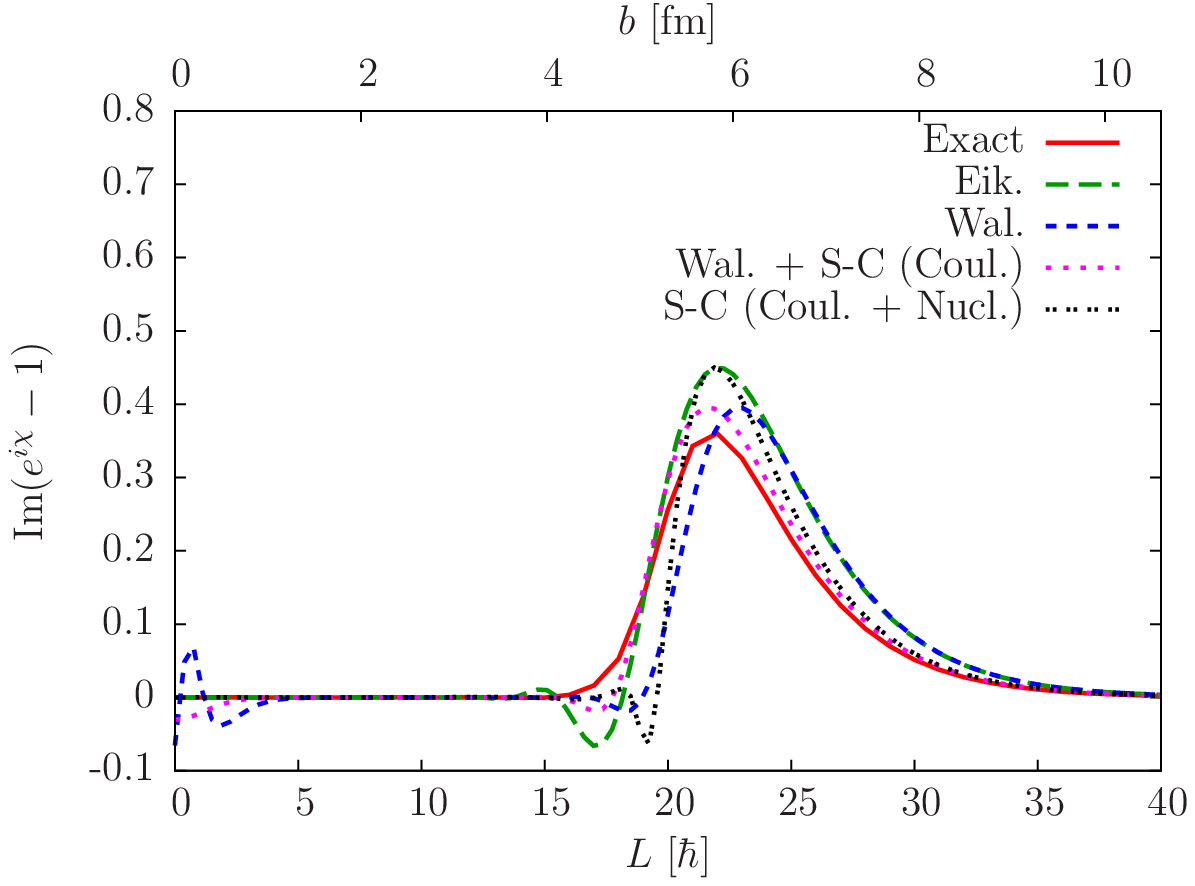}
	}}
	\caption{Elastic scattering of $^{10}\mathrm{Be}$ off $^{12}\mathrm{C}$ at 67~MeV/nucleon (a,b) and 10~MeV/nucleon (c,d). The cross sections are plotted as a ratio to Rutherford as a function of the scattering angle $\theta$ (a,c) and the imaginary part of the $T$-matrices as a function of the angular momentum and the corresponding impact parameter (b,d). The results are obtained with the partial-wave expansion (red solid line), the standard eikonal approximation (green long dashed line), Wallace's correction (blue short dashed line, see \Sec{CorrWallace}), Wallace's correction combined with the semi-classical Coulomb correction~\Eq{EqCorrSCC} (magenta dotted line) and the semi-classical correction~\Eq{EqCorrSCNC} (black dotted line).}\label{Fig10Be67-10EikWalSC}
\end{figure*}

In Fig.~\ref{Fig10Be67-10EikWalSC}, the red solid line corresponds to this exact method, the green long dashed line to the standard eikonal approximation, the blue short dashed line to Wallace's correction (see \Sec{CorrWallace}), the magenta dotted line to Wallace's correction combined with the semi-classical Coulomb correction~\Eq{EqCorrSCC} and the black dotted line to the semi-classical correction~\Eq{EqCorrSCNC}.
We only plot the first order of Wallace's correction since the corrective terms of the second and the third orders are negligible. 

Wallace's correction leads to nearly exact results at high energy (67~MeV/nucleon in Figs.~\ref{Fig10Be67-10EikWalSC}(a) and~\ref{Fig10Be67-10EikWalSC}(b)) but is less effective at lower energy (10~MeV/nucleon in Figs.~\ref{Fig10Be67-10EikWalSC}(c) and~\ref{Fig10Be67-10EikWalSC}(d)) since there are still discrepancies with the exact results. Nevertheless at low energy, the oscillations pattern of the cross sections is better reproduced than with the standard eikonal approximation. But the correction induces a shift of the cross sections to more forward angles and of the $T$-matrices to larger impact parameters. It can be explained by the attractive feature of the nuclear interaction: as Wallace's correction aims to improve the simulation of this interaction within the eikonal model, it causes an underestimation of the scattering angle. 

To counter this shift, the Coulomb repulsion needs to be better accounted for. This motivates the introduction of the semi-classical Coulomb correction~\Eq{EqCorrSCC}, which leads to the results plotted magenta dotted line. At high energy, it has no impact and the good agreement is kept. At low energy, the semi-classical Coulomb correction compensates the shift induced by Wallace's correction and the calculations are now in phase with the exact ones.  However, the resulting cross sections still lie above the exact ones at large angles. Therefore, we should enhance the absorption at small impact parameters to increase the accuracy.

This need for higher absorption as well as the desire to have only one consistent correction to both interactions has driven us to analyse the semi-classical correction~\Eq{EqCorrSCNC}.  
At high energy, the accuracy of the eikonal model is worsened while at lower energies one can note a small improvement at forward angles (below 20$^\circ$). It also reproduces well the oscillations pattern of the cross sections without inducing any shift in the results. Yet, even at low energy, this correction is still insufficient at large angles due to a lack of absorption.

\section{Conclusions}\label{Conclusions}

The study of halo nuclei is performed through measurements of reactions processes. To properly infer informations about the structure of these exotic nuclei from the measurements, we need an accurate model to describe the three- and four-body collisions.  State-of-the-art models such as CDCC can become expensive from a computational point of view and can present some convergence issues at energies aimed by RIB facilities such as ISOLDE at CERN (10~MeV/nucleon). The eikonal model is simple and presents a reduced computational time. This has motivated the study of its extension to low energies. 

The present work is a first step towards a better description of three- and four-body collisions. We have analysed two corrections and their interplay in a simpler case, a two-body collision of light nuclei. Both corrections aim to account for the deflection of the projectile by the target due to both interactions, neglected in the standard eikonal model. To evaluate the accuracy gain brought by each corrections, we have computed the angular distribution of the cross sections for the elastic scattering of \ex{10}Be off \ex{12}C at two energies (67 and 10~MeV/nucleon).

The first correction was developed by Wallace and is derived from a perturbation development of the $T$-matrix around the eikonal propagator with a corrective term that improves the simulation of the nuclear interaction~\cite{Wallace}.  Since the convergence is fast, only the first order has been presented. The second correction relies on a semi-classical approach and acts only on the Coulomb interaction or on both Coulomb and nuclear interactions.

Results have shown that Wallace's correction is more efficient at high energies (67~MeV/nucleon) and reproduces well the oscillation pattern of the angular distribution. It also induces a shift to more forward angles at low energies (10~MeV/nucleon) which is cancelled when the semi-classical Coulomb correction is introduced~\cite{Bertulani,BrogliaWinther}.  Both corrections combined enable an extension of the eikonal approximation to lower energies but the cross sections are still overestimated at large angles. 

To cope with this inadequacy and to have one consistent correction, we have studied the semi-classical correction applied to both Coulomb and nuclear interactions~\cite{AguiarZardiVitturi1997,LenziVitturiZardi1995}. The analysis has pointed out that there are no significant accuracy gain at low energies and that at high energies, the eikonal model leads to less precise results. The only improvement is the reproduction of the amplitude of the oscillations. 

In conclusion, we have achieved an extension of the eikonal model to low energies but both corrections tested have failed at reproducing the absorption at large angles. To enhance absorption, we could apply the semi-classical correction~\Eq{EqCorrSCNC} with a complex distance of closest approach computed from the classical trajectory considering the whole optical potential~\cite{AguiarZardiVitturi1997}. Therefore, the imaginary part of the potential would be increased for small impact parameters, hopefully this would cause a reduction of the cross sections at large angles.

\nocite{*}

\begin{acknowledgements}
	C. Hebborn acknowledges the support of the Fund for Research Training in Industry and Agriculture (FRIA), Belgium. P. Capel acknowledges the support of the Deutsche Forschungsgesellschaft (DFG) with the Collaborative Research Center 1245. This work is part of the Belgian Research Initiative on eXotic nuclei (BriX), program P7/12
	on inter-university attraction poles of the Belgian Federal Science Policy Office.
	This project has also received funding from  from the European Union's Horizon 2020 research and innovation program under grant agreement No 654002.
\end{acknowledgements}

\end{document}